\documentclass[letter]{aa} 

\usepackage{natbib}
\usepackage{graphicx}
\usepackage{txfonts}
\begin{document}

   \title{The formation history of our Galaxy's nuclear stellar disc
     constrained from HST observations of  the Quintuplet field }
\authorrunning{Sch\"odel et al.}
\titlerunning{NSD formation from Quintuplet field}


  \author{R. Sch\"odel
          \inst{1}
          \and
          F. Nogueras-Lara
          \inst{2}
          \and
          M. Hosek Jr.
          \inst{3}
          \and
          T. Do
          \inst{3}
          \and
          J. Lu
          \inst{4}
          \and
          A. Mart\'inez Arranz
          \inst{1}
          \and
          A. Ghez
          \inst{3}
          \and
          R. M. Rich
          \inst{3}
          \and
          A. Gardini
          \inst{1}
          \and
          E. Gallego-Cano
          \inst{1}
         \and 
          M. Cano Gonz\'alez
          \inst{1}
         \and 
          A. T. Gallego-Calvente
          \inst{1}
         }
   \institute{
    Instituto de Astrof\'isica de Andaluc\'ia (CSIC),
     Glorieta de la Astronom\'ia s/n, 18008 Granada, Spain\\
              \email{rainer@iaa.es}
  \and
  European Southern Observatory, Karl-Schwarzschild-Strasse 2, D-85748 Garching bei M\"unchen, Germany  \\
   \and
   Department of Physics and Astronomy, University of California, Los Angeles, CA 90095, USA\\
   \and
   Department of Astronomy, 501 Campbell Hall, University of
   California, Berkeley, CA, 94720, USA\\
     }

   \date{;}

 
  \abstract
{ Until recently it was
     thought that the nuclear stellar disc at the centre of our Galaxy
     was formed via
     quasi-continuous star formation over billions of years. However,
     an analysis of GALACTICNUCLEUS survey data indicates
     that $>80$\% of the mass of the stellar disc formed at least
     8 Gyr ago and about $5\%$ roughly 1\,Gyr ago.}
   {Our aim is to derive new constraints on the formation history of the nuclear stellar
     disc.}
   {We analysed a catalogue of HST/WFC3-IR observations of the
     Quintuplet cluster field.  From this catalogue, we selected
       about $24000$ field stars that probably belong to the nuclear
       stellar disc. We used red clump giants to deredden the sample and fit
     the resulting F153M luminosity function with a linear combination
     of theoretical luminosity functions created from different
     stellar evolutionary models.}
   {We find that
     $\gtrsim$$70\%$ of the stellar population in the nuclear disc
     probably formed
     more than $10$\,Gyr ago, while
     $\sim$$15\%$ formed in an event (or  series of events)
     $\sim1$\,Gyr ago. Up to 10\% of the stars appear to have formed
     in the past tens to hundreds of Myr. These results 
     do not change significantly for reasonable variations in the
     assumed mean metallicity, sample selection, reddening
     correction, or stellar evolutionary models.}
 {We confirm previous work that changed the formation paradigm for
   stars in the Galactic Centre. The nuclear stellar disc is indeed a
   very old structure. There seems to have been little  star
   formation activity between its formation and about 1\,Gyr ago.}


   \maketitle
%

\section{Introduction}

The Galactic Centre is marked by the presence of the 
    nuclear stellar  disc (NSD), a dense, flat, rotating stellar structure of
     roughly one billion stars, aligned with the Galactic plane, with exponential scale lengths of
     about 90\,pc in the radial direction and 30\,pc in the vertical
     direction \citep{Sormani:2022dq}. It probably formed from gas
     transported to the centre through the Galactic bar, in a similar
     way to nuclear discs in other galaxies
     \citep{Gadotti:2020xq,Bittner:2020qx}.

     The Galactic Centre is observationally challenging \citep[see
     section 2 in][] {Schodel:2014bn}. Due to the extreme crowding of
     stars, studies of its stellar population
     should be carried out with angular resolutions $\lesssim0.2"$,
     that is, either from space, with the HST or JWST, or from the
     ground, with speckle or adaptive optics techniques. Due
     to the high interstellar extinction toward the Galactic Centre,
     sensitive observations are mostly restricted to the wavelength
     range $\gtrsim1.5\,\mu$m, where intrinsic stellar colours are
     small. Since extinction also varies on arcsecond scales, it is
     extremely difficult to use colours to distinguish between
     different types of stars, but the observed colours can serve to deredden
     the observed stellar magnitudes. As a result, studies of the star
     formation history of the Galactic Centre either use high angular
     resolution integral field spectroscopy to investigate small
     regions, such as the central parsec \cite [e.g.][]{Pfuhl:2011uq,Chen:2022aw}, or they rely on the analysis of luminosity
     functions, when studying large areas photometrically
     \citep{Figer:2004fk,Nogueras-Lara:2020pp,Schodel:2020qc}.

     \citet{Figer:2004fk} carried out the first study to probe the
     star formation history of the Galactic Centre across several
     widely dispersed but small fields with HST/NICMOS, and used   Gemini adaptive optics data
     (and also  seeing-limited Lick photometry) to analyse luminosity
     functions. They found the data to be consistent with continuous star formation at an average rate
     of $0.01\,M_{\odot}$yr$^{-1}$. They did not distinguish between
     different structures at the Galactic Centre, such as the nuclear star
     cluster, the NSD, and the inner bar region,
     probably because our knowledge about these different structures
     was still very limited at the time.

     The next investigation of the star formation history of the
     Galactic Centre on large scales, specifically of the NSD was carried out by
     \citet{Nogueras-Lara:2020pp}. They used $0.2"$ angular resolution
     data from the GALACTICNUCLEUS survey,  the most complete
     survey of the Galactic Centre region
to date     \citep{Nogueras-Lara:2019yj}. They studied a
     $90$\,pc$\times30$\,pc field, centred on Sagittarius\,A*, but
     omitted the nuclear star cluster due to its excessive
     crowding. They found that $\gtrsim80\%$ of the stellar mass
     formed at least 8\,Gyr ago, followed by a long quiescent period
     that was ended by a strong star formation event $\sim$1\,Gyr ago,
     when about 5\% of the stellar mass formed. Up to a few precent of the
     stars formed in the past few 100\,Myr.

     In this work we follow up the work by
     \citet{Nogueras-Lara:2020pp} and study the star formation history
     in the NSD with a fully independent data set. We chose to use HST
     WFC3-IR data on a
     $\sim$$2'\times2'$ field centred on the Quintuplet young massive
     cluster, which is located at a projected distance of about 30\,pc
     from the central black hole, Sagittarius\,A*, very nearly on the
     Galactic plane: (l,b) = (0.164, -0.0602) deg. The field is
     located far outside the half-light radius of the nuclear star
     cluster
     \citep[$\sim$4\,pc][]{Schodel:2014fk,Feldmeier-Krause:2017rt}, so
     that we do not have to worry about contamination of the stellar
     population from the nuclear cluster \citet{Rui:2019ch} had
     reported on the presence of  a potential secondary red clump (RC) in the
     colour-magnitude diagram (CMD) of this region, similar to what
     was reported by \citet{Nogueras-Lara:2020pp} in their study of
     the star formation history of the NSD.  \citet{Hosek:2022om}
     published a catalogue containing $\sim$$40,000$ stars in this
     field, with $F153M$, $F139M$, and $F127M$ magnitudes and
     precision proper motions. The kinematic information allows us to
     clean the data from young stars that belong to the Quintuplet
     cluster and also to minimise pollution from interloping inner
     Galactic bar stars.


\section{Data}

\begin{figure}[!htb]
\includegraphics[width=\columnwidth,angle=0]{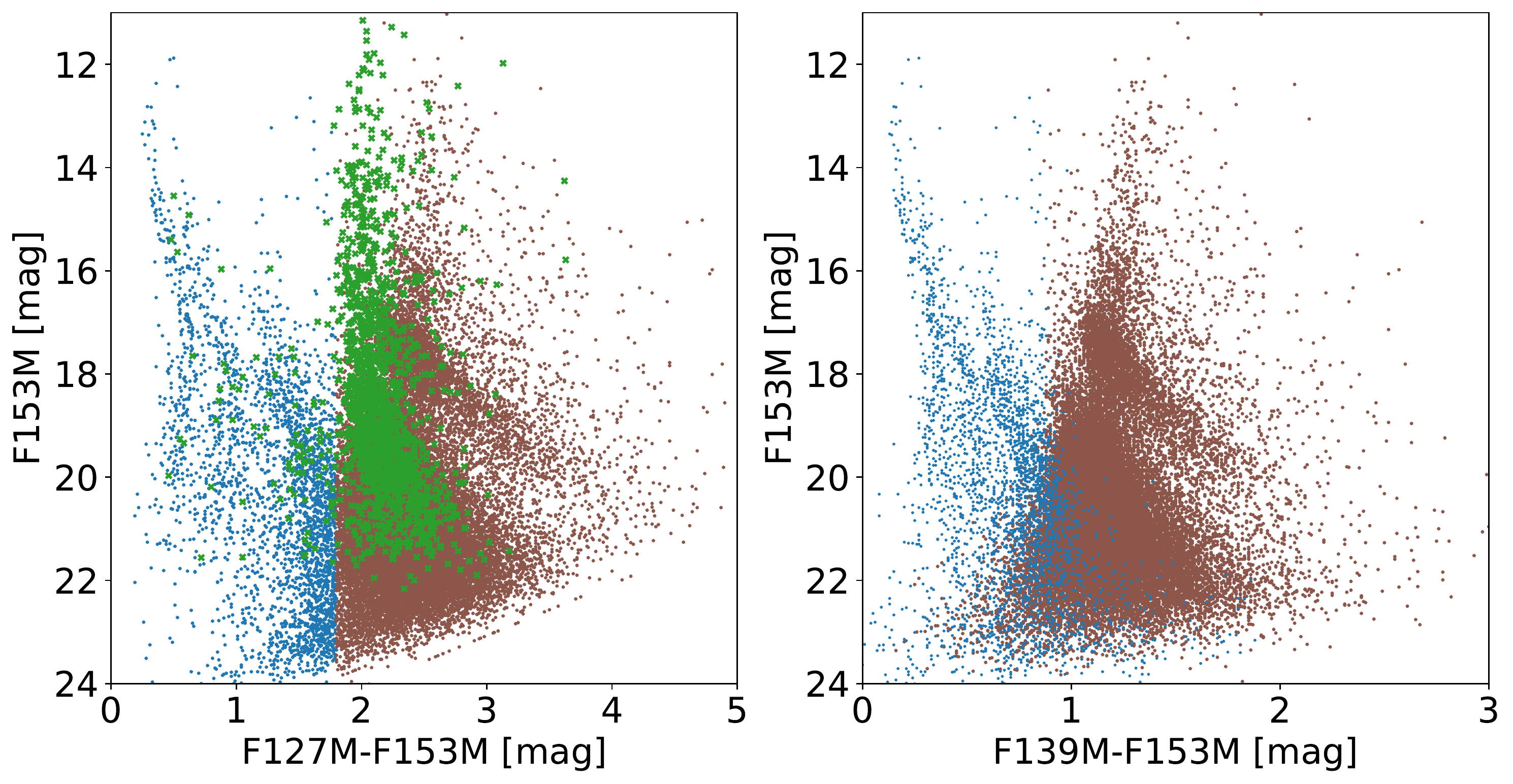}
\caption{\label{Fig:CMD_all} Colour-magnitude diagrams for the Quintuplet
  field. Foreground stars
  ($F127M-F153M < 1.8$\,mag) are shown in blue, Galactic Centre stars
  in brown, and probable Quintuplet cluster members (membership
  probability from proper motions $>50\%$) as green crosses.  Some
  foreground and background stars move along with the bulk of the
  Quintuplet and are therefore also shown as      green crosses \citep[see][]{Rui:2019ch}.  The probable
  members of the Quintuplet cluster are only shown in the left panel.}
\end{figure}

\begin{figure}[!htb]
\includegraphics[width=\columnwidth,angle=0]{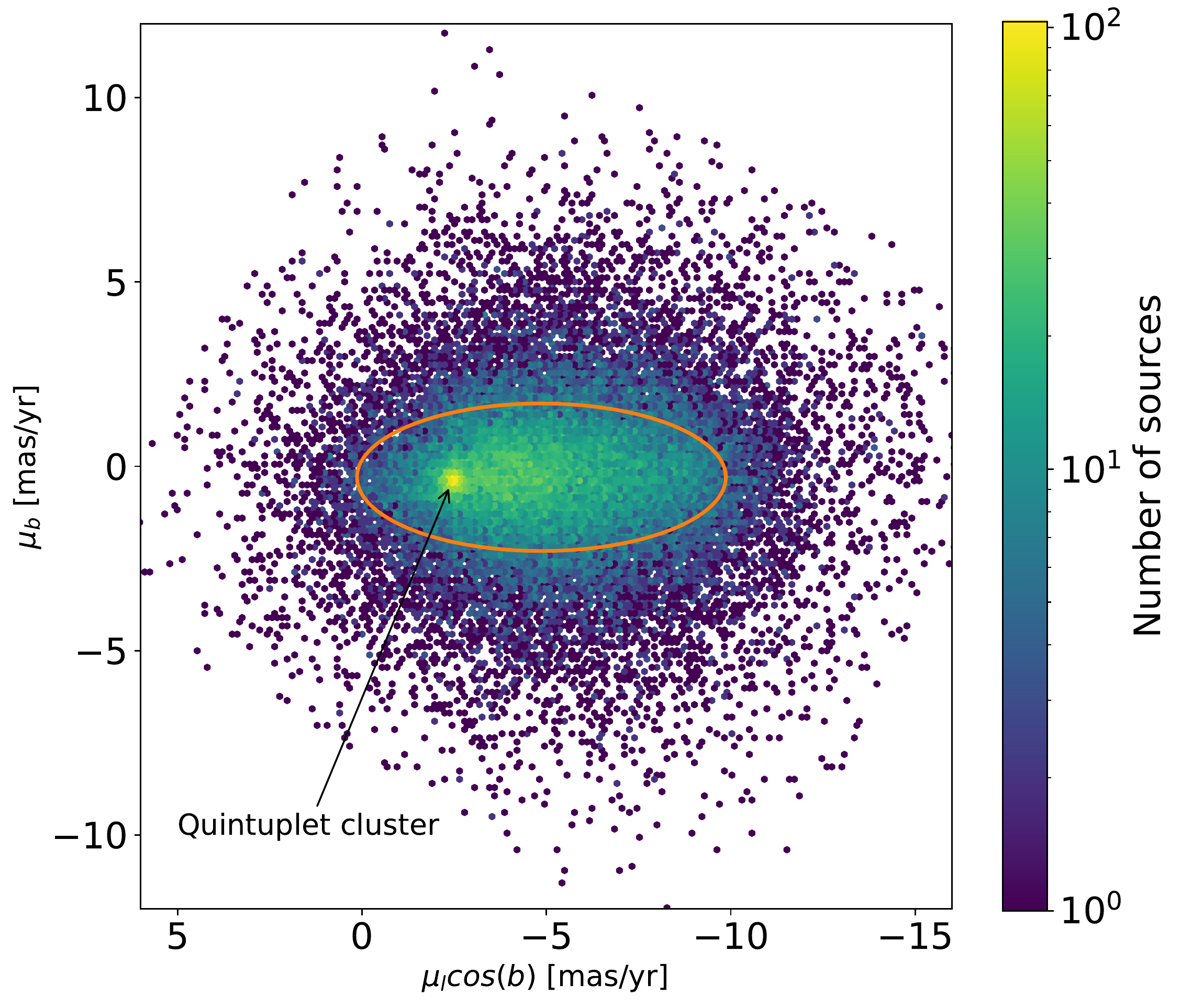}
\caption{\label{Fig:VPD} Vector point diagram for the catalogue for
  the Quintuplet field. The quality cuts of \citet{Hosek:2022om} have
  been applied to include only high-quality proper motions in this
  figure.  Shown are the number of sources per proper
    motion bin, considering the velocities  parallel and perpendicular to the
    Galactic Plane, $\mu_{b}$ and $\mu_{l}cos(b)$, respectively.  The arrow points to
  the position of the Quintuplet cluster in proper motion space. The orange ellipse indicates our selection
  criterion for nuclear stellar disc members.}
\end{figure}

We use the catalogue from \citet{Hosek:2022om}, but without any cuts
applied to proper motion uncertainties so as to increase completeness
at the faint end of the luminosity function.  We show the
corresponding $F153M$ versus $F127M-F153M$  CMD in Fig.\,\ref{Fig:CMD_all}. The CMD is rich in features and shows
the foreground stellar population (blue), stars in the Quintuplet
cluster (green), and stars that belong to the surrounding Galactic Centre
population (brown). In the following we  use term  field
  population for the Galactic Centre population
that does not belong to the Quintuplet.  The proper motion vector point
diagram in Fig.\,\ref{Fig:VPD} shows the flattened shape expected for
the NSD, surrounded by a more spherical distribution that corresponds
to the stars from the inner bar \citep{Shahzamanian:2022vz}. Finally, the Quintuplet cluster
stands out as a clear overdensity in proper motion space. 

In the brightness ranges considered here, foreground stars towards the
Galactic Centre can be identified readily via their low reddening \citep[e.g.][]{Nogueras-Lara:2021cf}.
We preprocessed the data by excluding foreground stars
($F127M-F153M<1.8$\,mag), background stars (stars with very high
reddening: $F127M-F153M>2.8$\,mag),
and members of the young Quintuplet cluster (membership probability
$>50\%$ according to the catalogue). The remaining field population displays the prominent RC feature, located around $F153m\approx17.5$ and
$F139M-153M\approx1.3$ and elongated along the reddening vector (right
panel of Fig.\,\ref{Fig:CMD_all}).

We   studied the luminosity function in the $F153M$ filter, where
extinction is least severe for the filters used here and the star
counts are consequently the deepest. Since differential reddening
washes out the features in the luminosity function, we first
dereddened the data. The intrinsic colours of
stars in the RC  are small and very similar, and are
largely independent of metallicity (at least between solar and twice
solar) and age (for ages $\gtrsim2$\,Gyr) \citep{Girardi:2016fk}.

\begin{figure}[!htb]
\includegraphics[width=\columnwidth,angle=0]{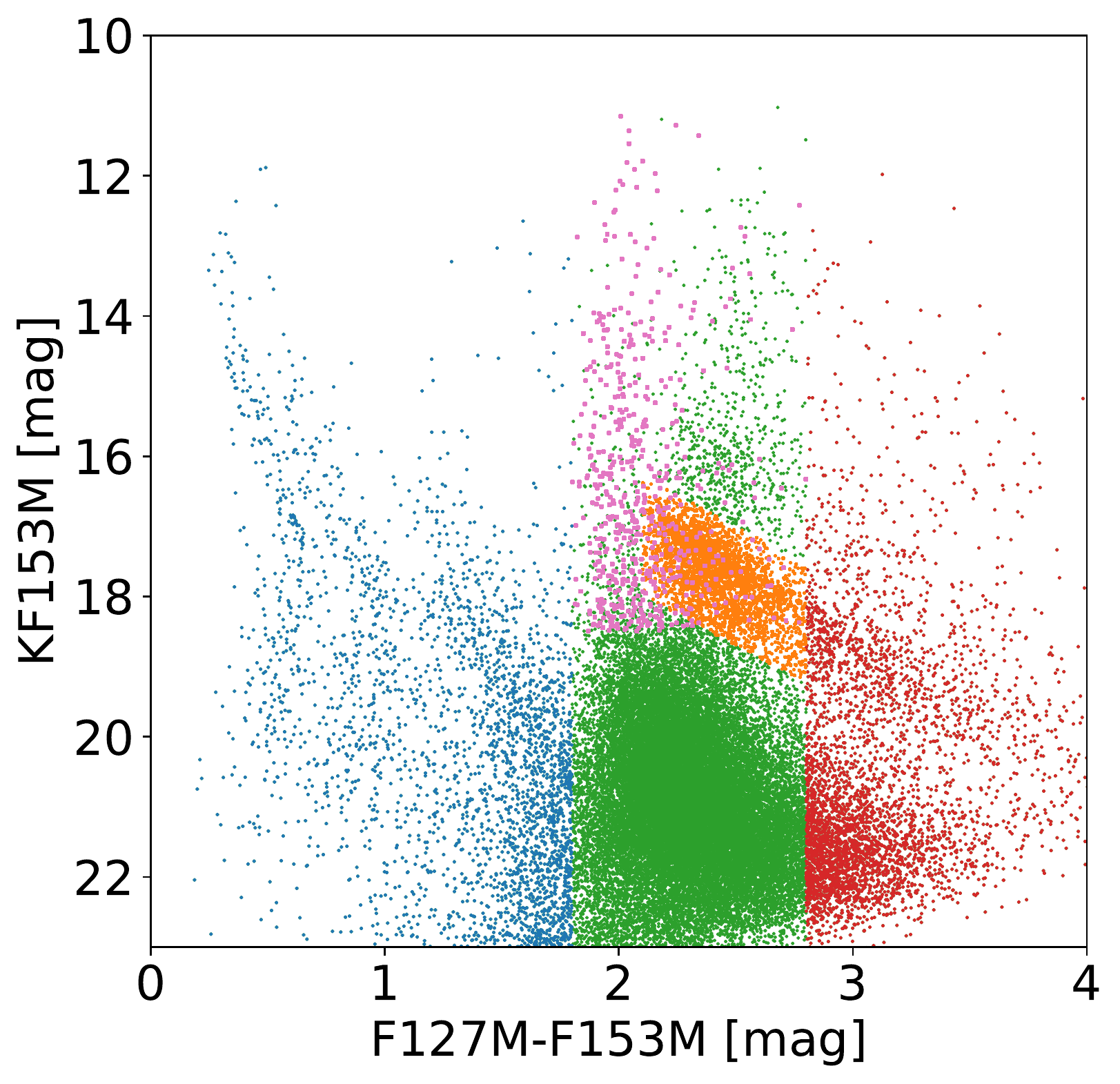}
\caption{\label{Fig:CMD_types} Colour-magnitude diagram. Blue:
  Foreground stars; red: background stars; orange: RC stars used for dereddening; pink:
  young stars from the Quintuplet cluster. }
\end{figure}

We therefore chose RC stars as  dereddening stars (see Fig.\,\ref{Fig:CMD_types}) and assumed
intrinsic colours of $F127M-F139M=0.18$, $F139M-F153M=0.2$, and
$F127M-F153M=0.4$ \citep[from
MIST\footnote{http://waps.cfa.harvard.edu/MIST/interp\_isos.html,
  version 1.2} isochrones of metallicity $Z=0.03$ and
age 3\,Gyr;][]{Dotter:2016sh,Choi:2016qh,Paxton:2011vh,Paxton:2018tk}. We assumed an extinction
curve of $A_{\lambda}\propto\lambda^{-\alpha}$ with $\alpha = 2.1$
\citep{Fritz:2011fk}.  We computed the
mean of the extinction values that resulted from the three different possible filter combinations.

Each star in the catalogue was dereddened when there were at least
three dereddening stars within a radius of $2.5"$. We used the median
of the extinction values corresponding to the dereddening stars. Similar dereddening methods were applied  by
\citet{Schodel:2010fk} and \citet{Nogueras-Lara:2021di}, among others. Due to the
distribution of stars along the line of sight and the small-scale
variable extinction towards the Galactic Centre,  this dereddening method will
produce some outliers. 
After dereddening, we deselected the outliers by excluding all stars
$>2.5\,\sigma$ to the red or blue side of the median of the
dereddened $F139M-F153M$ CMD, along the reddening line.

\begin{figure}[!htb]
\includegraphics[width=\columnwidth]{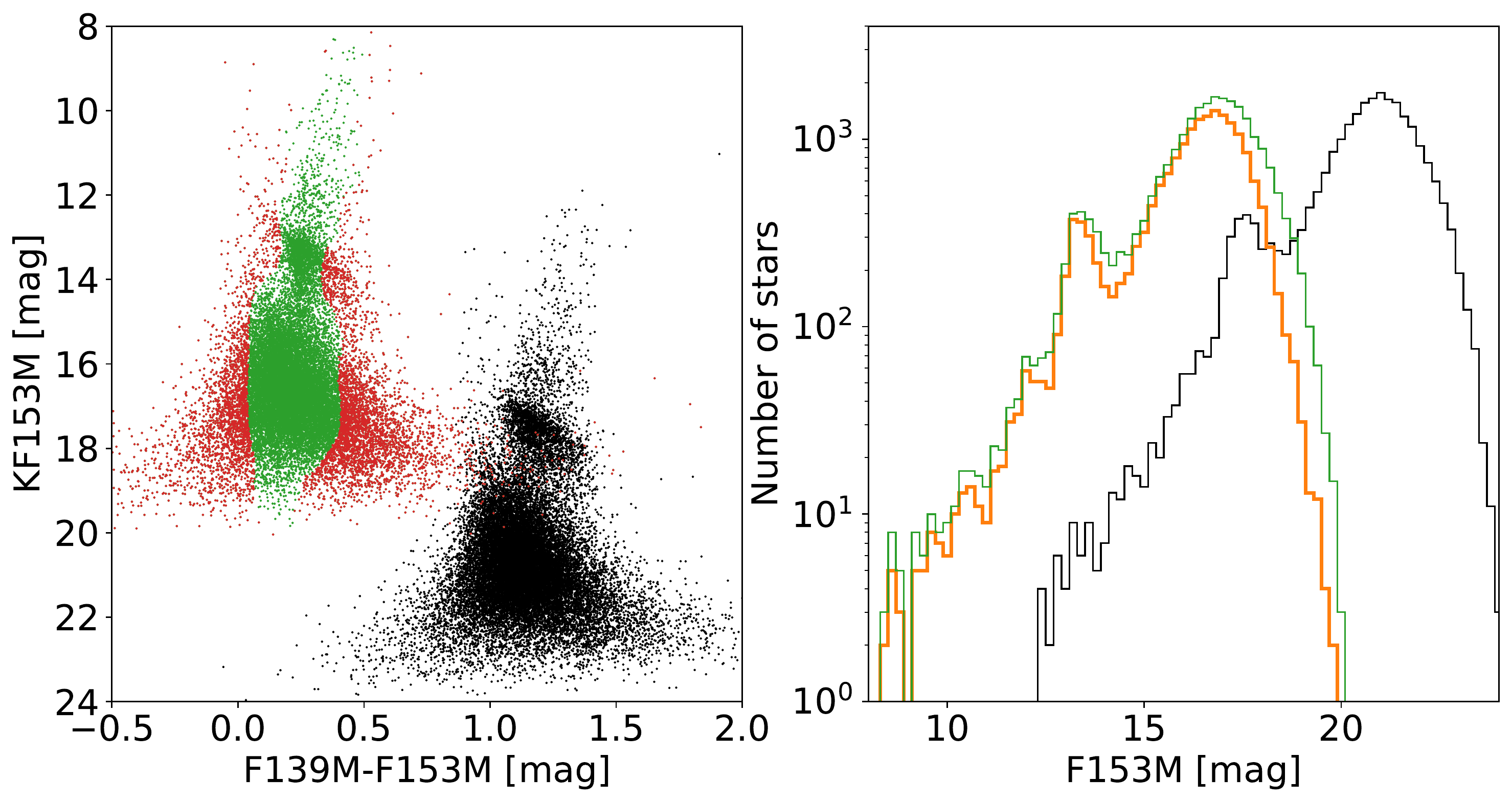}
\caption{\label{Fig:dered} Observed (black) and
  dereddened (blue) CMDs and luminosity functions. Excluded outliers
  are shown in red in the CMD. At faint magnitudes the green
  `cloud' becomes narrower because of the systematics of the
  outlier estimation (red sources drop out). This does not have any
  influence on our analysis, because these faint stars lie several
  magnitudes below our completeness cut-off. The orange luminosity function
  excludes the  outliers after dereddening. }
\end{figure}

 The original catalogue had $46087$
stars detected in $F153M$. After applying  all the  above-described selection criteria, we were
left with $24017$  stars. In Fig.\,\ref{Fig:dered} we show the measured
and dereddened CMDs as well as the measured and dereddend luminosity
functions.

Nuclear stellar disc stars and stars from the innermost Galactic bar will occupy the
same space, although the fraction of the latter will be below 25\% at
the location of the Quintuplet \citep[see Fig.\,10
in][]{Sormani:2022dq}.  According to the analysis of
\citet{Shahzamanian:2022vz} the velocity dispersion of stars in the
NSD is $\sigma_{\perp}=1.5$\,mas\,yr$^{-1}$
($\sigma_{\parallel}=2.0$\,mas\,yr$^{-1}$) in the direction
perpendicular (parallel) to the Galactic plane. The velocity
dispersion of bar stars is close to isotropic with
$\sigma_{Bar}\approx3$\,mas\,yr$^{-1}$.  We suppress any potential
contamination by bar stars further by selection in velocity
space. We chose as NSD sample the stars inside an ellipse centred on
the mean proper motions, with a half width of $5.0$\,mas\,yr$^{-1}$
and and a half height of $2.0$\,mas\,yr$^{-1}$, parallel and
perpendicular to the Galactic plane, respectively (orange ellipse in
Fig.\,\ref{Fig:VPD}). As can be seen in Fig.\,9 of
\citet{Shahzamanian:2022vz}, we  thus cut off the high-velocity
tail of the bar stars. Our criterion is admittedly somewhat
arbitrary, but our results are hardly sensitive to the exact choice of
the selection criterion in velocity space (see discussion of
systematics below).

{\it Completeness.} At the angular resolution of the used
observations, completeness is dominated by crowding. We used the {\it
  critical distance} method \citep[see
e.g.][]{Eisenhauer:1998tg,Harayama:2008ph}. We found that
completeness was close to 100\% up to an observed magnitude of
$F153M=17$, dropped to 76\% at $F153M=20$ and to 42\% at $F153M=21$.
We limited our analysis to stars brighter than an observed magnitude
of $F153M=20.5$ (about $F153M=15.5$ after extinction correction) and
applied the estimated completeness correction to all star counts,
assuming a 5\% uncertainty in the estimated completeness
levels. Since we are limited by crowding and the observations are
sufficiently sensitive, the completeness at $F127M$ is almost
identical to that at $F153M$, which can be seen in the CMD in the
left panel of Fig.\,\ref{Fig:CMD_all}, which does not show any
colour-dependent cut-off at faint magnitudes.

\section{Star formation history of the NSD}

\begin{figure}[!htb]
\includegraphics[width=\columnwidth]{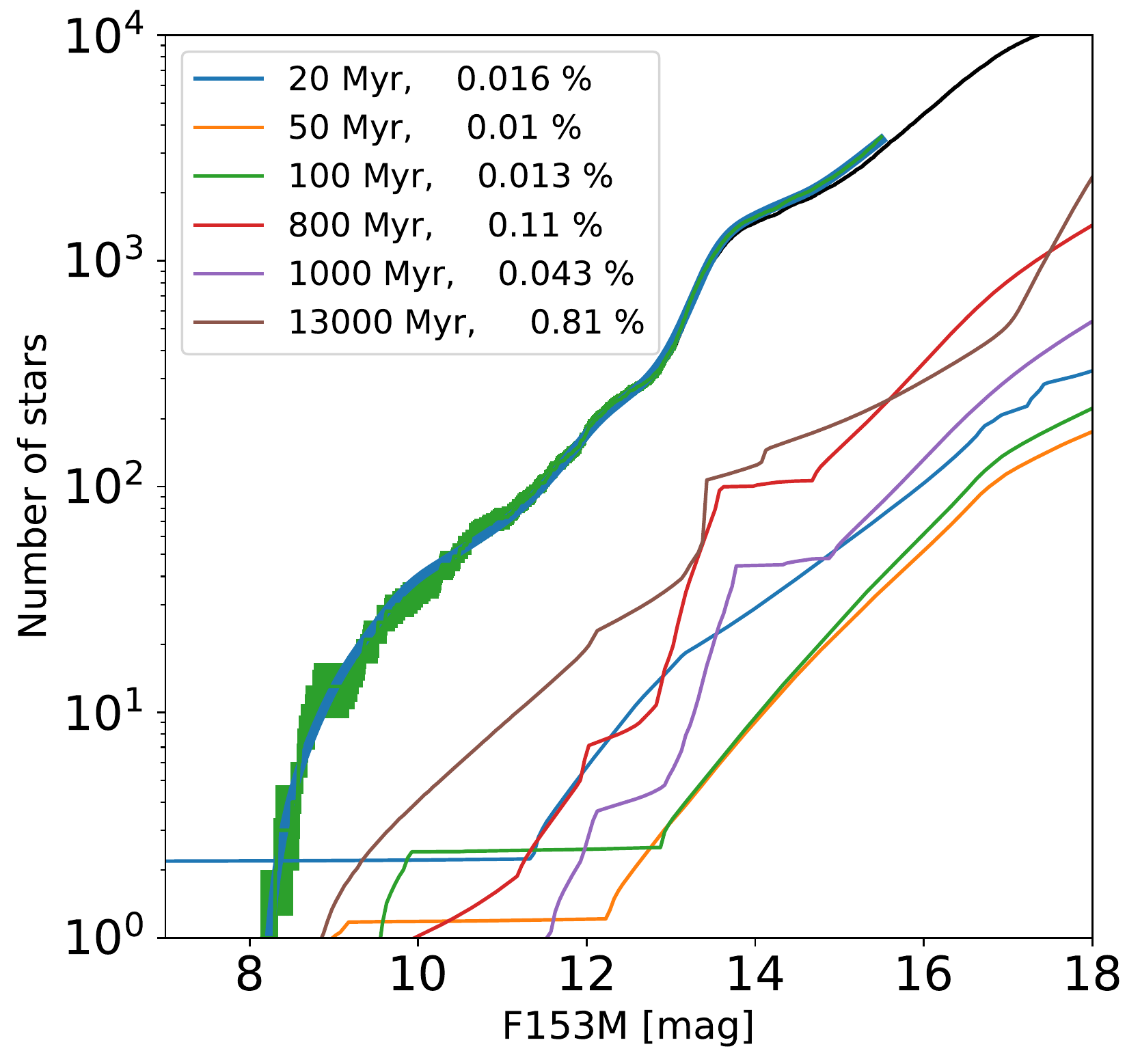}
\caption{\label{Fig:best-mist} Fit of cumulative theoretical
  luminosity functions to the measured cumulative luminosity function
  of the NSD (data: black; data after completeness correction and
  including error bars: green; best fit: blue). Theoretical functions
  of all stellar population that contribute more than 1\% of
  originally formed stellar mass are    colour-coded (see inset).}
\end{figure}

We fitted the cumulative luminosity function of our sample with
theoretical luminosity functions. Using the cumulative function avoids
biases due to the choice of the bin width. We assumed a linear
combination of 19 ages: 20, 50, 100, 200, 400, 600, 800, 1000, 1500,
2000, 3000, 4000, 5000, 6000, 7000, 8000, 9000, 11000, and 13000
Myr to cover the full range of expected ages. The luminosity functions
vary more rapidly in time for younger ages, which is the reason why we
chose a tighter spacing of ages below 3\,Gyr. The weight of all ages was initially set to zero. The fit
converged well and experiments with random starting values showed that
the fit (minimum $\chi^{2}$ with the Levenberg-Marquardt gradient
descent method) did apparently not get stuck in relative minima. The
theoretical luminosity functions were created from Basti\footnote{http://basti-iac.oa-abruzzo.inaf.it/index.html}
\citep[][]{Hidalgo:2018mw,Pietrinferni:2021bt}, PARSEC\footnote{http://stev.oapd.inaf.it/cgi-bin/cmd, release
1.25}\citep[][]{Bressan:2012xy,Chen:2014nr,Tang:2014rm,Marigo:2017fb,Pastorelli:2020pt},
and MIST isochrones,
assuming a solar-scaled metallicity of $Z=0.03$
\citep{Schultheis:2021du,Nogueras-Lara:2022by} and a Salpeter initial
mass function. A different choice of initial mass function had no
discernible effect because our sample consists almost entirely
of giant stars in a narrow mass interval of about $1-2\,M_{\odot}$
\citep[see also discussion in][]{Schodel:2020qc}. Additional
parameters to optimise the fit were residual extinction and a
Gaussian smoothing parameter \citep[to take into account effects such as measurement
uncertainties, remnant scatter from the imperfect reddening
correction, the line-of-sight extent of the NSD, or luminosity shifts
between the theoretical models; see][]{Schodel:2020qc}. The best fit for MIST isochrones is
shown in Fig.\,\ref{Fig:best-mist}. We limited the fit  to dereddened magnitudes
$F153M\leq15.5$, where completeness is $\gtrsim70\%$.

\begin{figure}[!htb]
\includegraphics[width=\columnwidth]{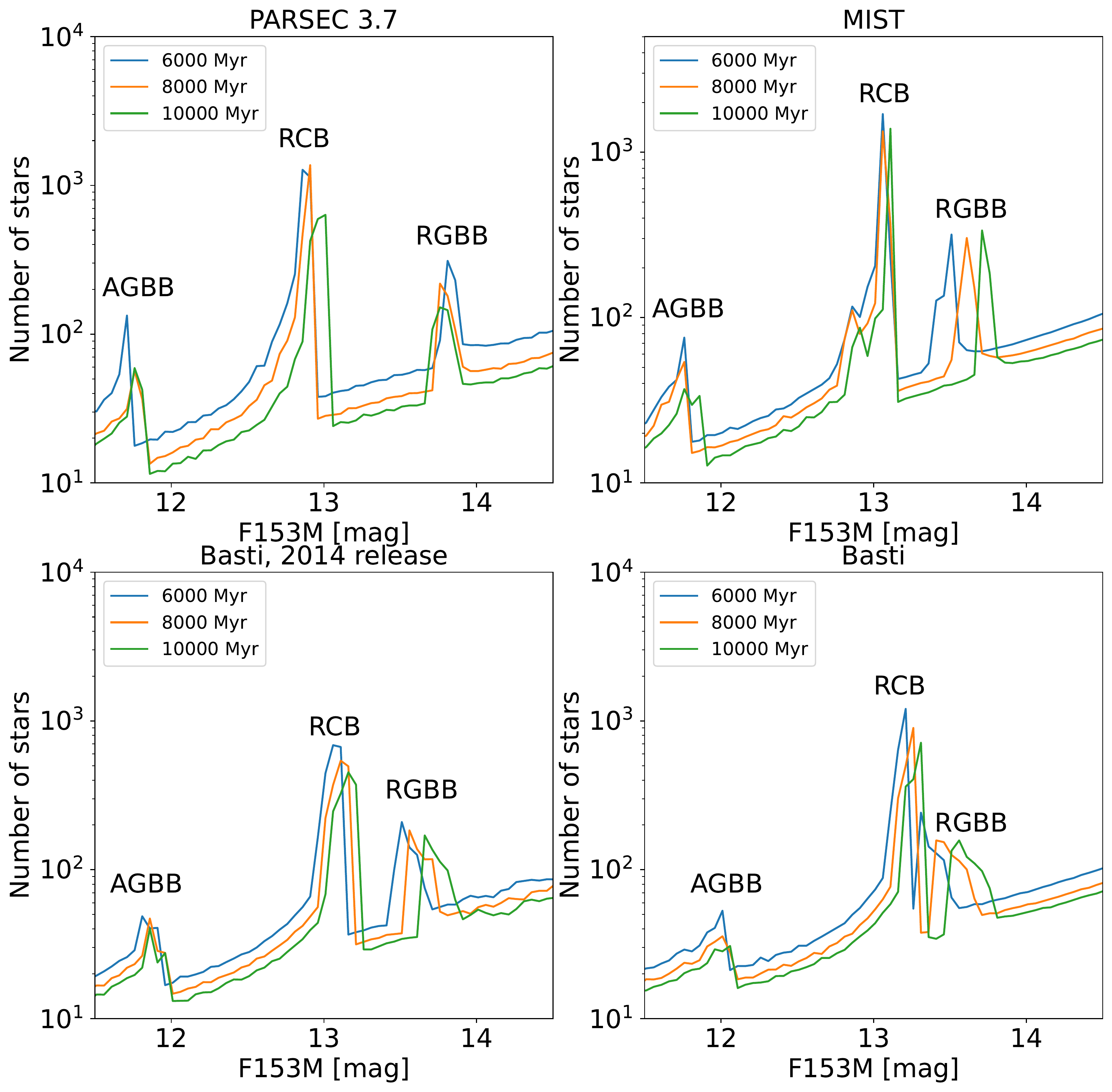}
\caption{\label{Fig:models} Comparison between theoretical luminosity
functions in the region around the asymptotic giant branch bump
(AGBB), RC bump (RCB) and the red giant branch bump (RGBB) for the MIST,
PARSEC, and Basti (current and 2014 release) evolutionary codes. The
AGB, RC, and RGB bumps are key features used to constrain the star
formation history from luminosity functions.  The stellar magnitudes assume no
extinction, but a distance modulus of $14.51$\,mag.}
\end{figure}

Since our data are limited in brightness, the main
features in the luminosity function that guide the convergence of the fits are the
cut-off at bright magnitudes, the asymptotic giant branch (AGB) bump, the RC bump (RCB), and the
red giant branch bump (RGBB). All
models differ in the separation between the RCB and
RGBB and its evolution with age. In Fig.\,\ref{Fig:models} we show the luminosity functions
between the AGB and the RGBB for PARSEC, MIST, and the new and old
(2014) Basti releases. PARSEC, MIST, and the 2014 release of Basti
models appear generally similar. Differences are apparent in the
magnitudes, shape, and time evolution of the RCB and RGBB peaks. The evolution of the
RGBB is inverse in PARSEC, as compared to MIST and Basti, that is it
becomes brighter with age instead of fainter. The new Basti release produces
markedly different luminosity functions than the other three models. The
RGBBs lie much closer to the RCB than in the other models and the RCB is
shifted to fainter magnitudes than in the other three models. Because
of this disagreement, we chose to use the old Basti release in our
analysis, but also show the Monte Carlo results (see below) for the
newer model.

\subsection{Monte Carlo simulations}

\begin{figure}[!htb]
\includegraphics[width=\columnwidth]{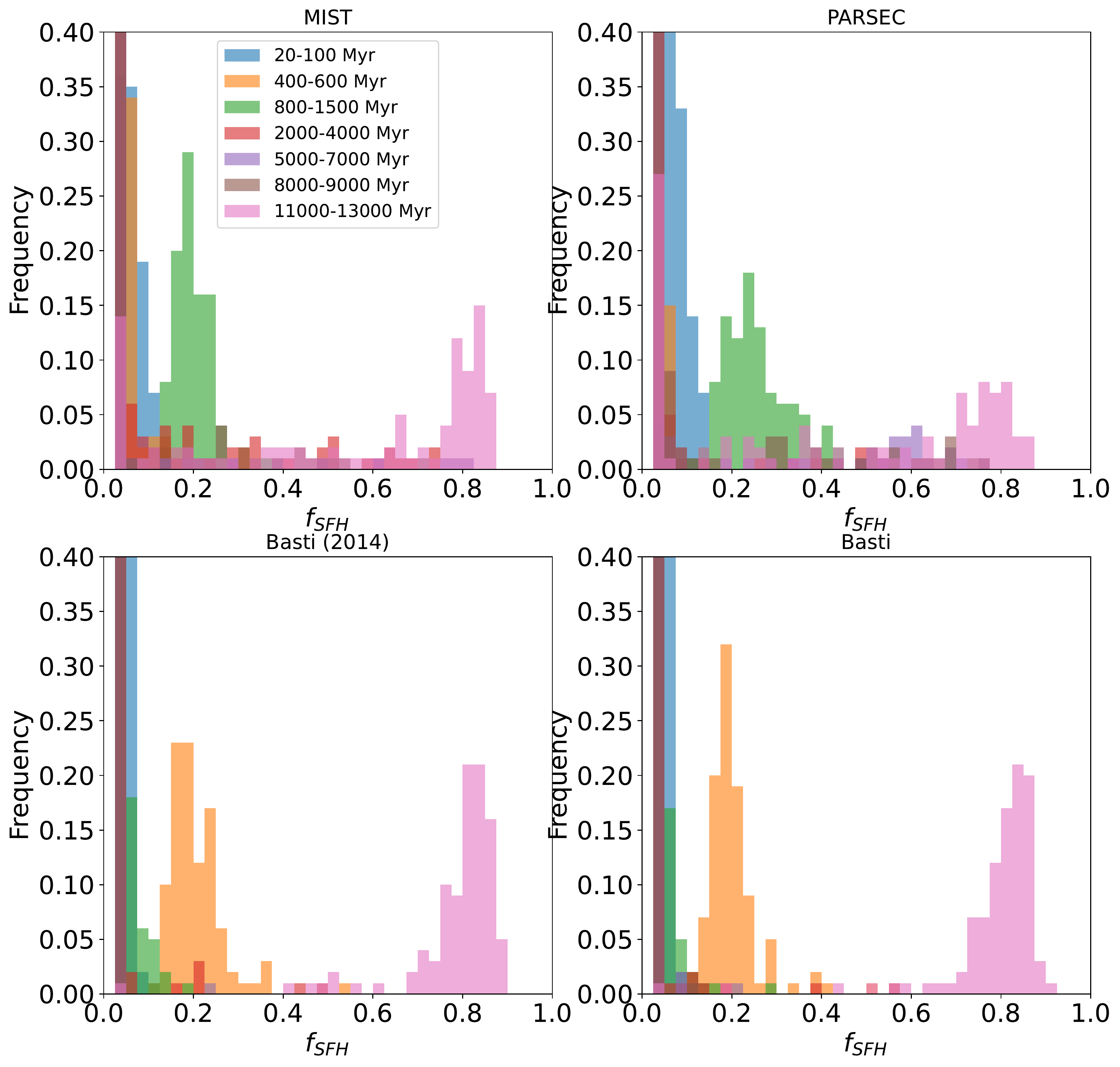}
\caption{\label{Fig:SFH-MC} Observed frequencies of different age bins
in the MC simulation of the star formation history. The different
panels correspond to different stellar evolution models, as labelled.}
\end{figure}

To constrain the distribution of possible solutions we carried out
Monte Carlo (MC) simulations with full sample bootstrap resampling of the
selected NSD stars and 100
iterations. The frequency histograms are shown in
Fig.\,\ref{Fig:SFH-MC}. The results of all simulations are similar,
but show some differences with respect to the underlying assumed
stellar evolution model. In all cases there are two clear peaks;  
the first  indicates that $\sim$80\% of star formation
took place $>10$\,Gyr ago and the second  shows that up to 20\% of the stellar mass formed around
0.8 to 1.5\,Gyr (MIST, PARSEC) or 0.4--0.6\,Gyr (Basti) ago.
There are no unambiguous signs of activity at
intermediate ages. Up to 10\% of the stars may have formed in
the past 100\,Myr.

\subsection{Best fits and their systematics}

In addition to the MC modelling, we inferred the best fit star
formation histories under different assumptions and checked the following sources of systematic uncertainty: 
\begin{enumerate}
\item Theoretical models: We compared the Basti (2014 release), MIST,
  and PARSEC models. 

\item Cut-off of the luminosity function at the faint end: The
  luminosity function becomes incomplete at faint magnitudes. The faint
  magnitude cut-off in the fitting of the luminosity functions can be
  important because, on the one hand, incomplete star counts may
  introduce a bias and, on the other hand, a fainter magnitude
  cut-off provides better constraints on the main sequence turn-off of
  certain populations. We fit  the models for 
  faint cut-offs at  $F153M=15.5$ (the default) as well as $F153M=15.0$  and
 $F153M=16.0$\,mag (extinction corrected magnitudes).

\item Metallicity: $1.5$ times solar (default), solar, and twice solar. 

\item Outlier rejection after dereddening: We performed the fit for
  3\,$\sigma$, 2.5\,$\sigma$ (default), and $2.0$\,$\sigma$ clipping.

\item Foreground star rejection: In addition to the default value
  $F127M-F153M =1.8$\,mag, we also used 1.7 and 1.9\,mag.

\item Background star rejection: In addition to the default value
  $F127M-F153M =2.8$\,mag, we also tested $2.6$ and $3.0$\,mag.

\item Different dereddening radii:  $1.5", 2.5"$, and $3.5"$.

\item Different rejection probabilities for Quintuplet cluster stars:
  $0.5$ (default), $0.3,$ and $0.7$.
\item Different selection of NSD stars: using, on the one hand, the
  stars inside an ellipse centred on the mean proper motions with a
  half width of $4.0$\,mas\,yr$^{-1}$ and a half height of
  $1.5$\,mas\,yr$^{-1}$, and, on the other hand,  a half-width of
  $6.0$\,mas\,yr$^{-1}$ and a half-height of
  $2.5$\,mas\,yr$^{-1}$. We also ran the fit using no velocity
  selection criterion, finding that this criterion had a negligible
  effect on our model fitting.

\item Binarity: The PARSEC models include Kroupa initial mass
  functions that are corrected for binarity. We used those
  binary-corrected luminosity functions. We also used the SPISEA code
  \citep{Hosek:2020mi} to prepare MIST luminosity functions that
  include unresolved binaries. We did not note any significant
  differences of the results with respect to the fits that do not take
  binarity into account.
\end{enumerate}

\begin{figure}[!htb]
\includegraphics[width=\columnwidth]{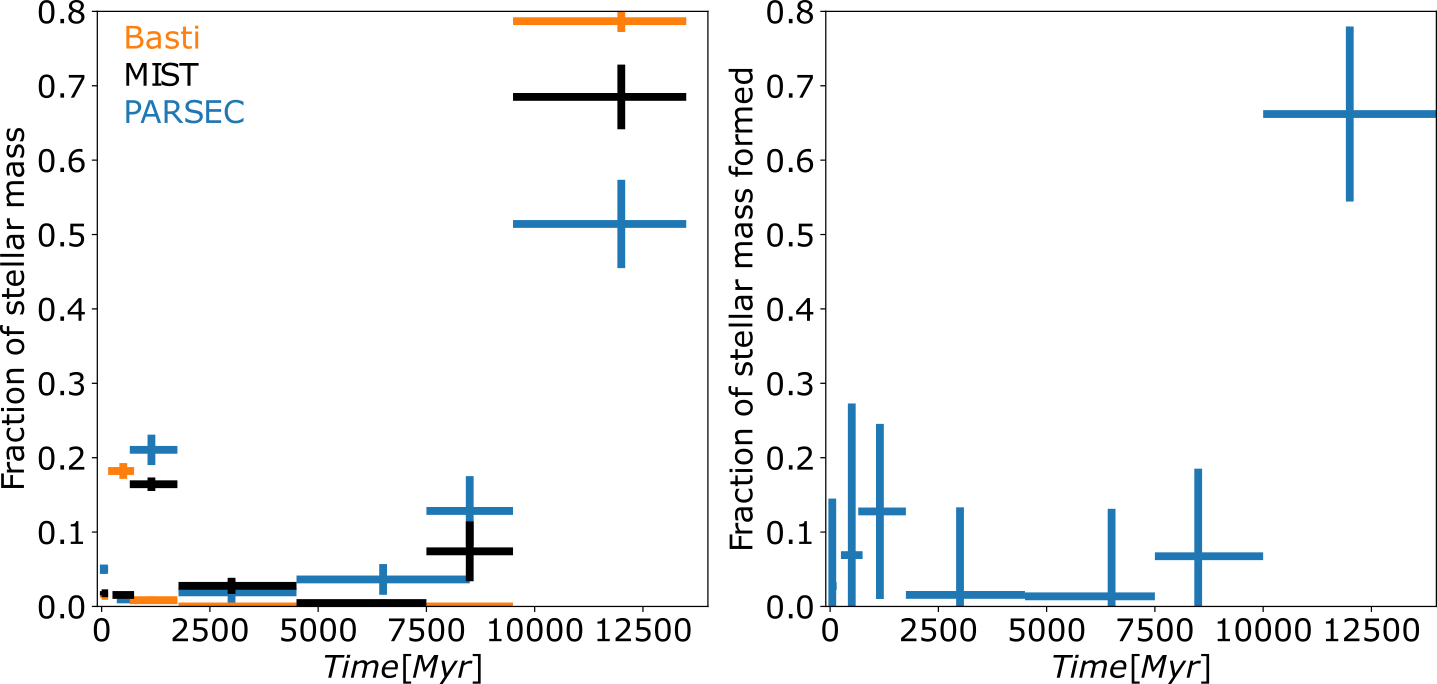}
\caption{\label{Fig:SFH-final} Star formation history of the NSD based
  on the mean and standard deviations of the best fits to the
  cumulative $F153M$ luminosity function, testing different
  sources of systematic uncertainties (see text).  Left:
  Resulting star formation history for different evolutionary models (blue: PARSEC; black:
MIST; orange: Basti). Right: Mean star formation history across all models and systematic assumptions.}
\end{figure}

We found the clearest differences between the different stellar evolutionary codes, as can be seen in
the left panel of Fig.\,\ref{Fig:SFH-final}, that shows means and
uncertainties across all assumptions, but shows those separately for
the different evolutionary codes. In particular, PARSEC
models tended to result in more solutions with some presence of star
formation at intermediate ages, as can also be seen in the MC
simulations (Fig.\,\ref{Fig:SFH-MC}).  Basti models also show an
age-shift compared to MIST and PARSEC: The strong star formation event
that appears  0.8-1.5\,Gyr for the latter two models appears at
0.6-0.8\,Gyr in the Basti models \citep[see also ][]{Nogueras-Lara:2020pp}. The right panel of
Fig.\,\ref{Fig:SFH-final} provides an overview of the overall star formation history,
averaged over different evolutionary models and systematic
assumptions. The result of our investigation of systematics is consistent with
the results from the MC simulations shown in Fig.\,\ref{Fig:SFH-MC}. 

\subsection{Star formation history in the Quintuplet field}

We find the following star formation history for the field population in the Quintuplet
field. About two-thirds of the stars formed at an age of
$11-13$\,Gyr. Star formation activity was very low between 2 and 10
Gyr ago. There was a clear
phase of very high activity about 1\,Gyr ago, when about 15\% of the
total stellar mass formed. Up to 10\% of the stellar mass may have
formed in the past few 100\,Myr.

\section{Discussion and conclusions}

In this work we followed up the secondary RC feature in the Quintuplet
field that was reported and discussed, but not investigated in depth,
by \citet{Rui:2019ch}. A double RC feature was also  reported for the entire
inner Galactic Centre region by \citet{Nogueras-Lara:2020pp} and interpreted
compellingly as a sign of intense star formation roughly 1\,Gyr ago. We
fit the $F153M$ luminosity functions of the Quintuplet field with
theoretical models, based on the PARSEC, Basti, and MIST evolutionary
codes.  Both the MC simulations and the repeated best fits with a broad range of
assumptions on systematic factors such as extinction correction,
metallicity, or faint luminosity cut-off, provide us with a consistent
picture: at least two-thirds  of the stars formed $>10$\,Gyr ago. This
initial phase was followed by a period of negligible star formation
until about 2\,Gyr ago. Subsequently, about 800\,Myr to 1\,Gyr ago, of
the order of 15\% of the stellar mass in this region were formed,
while up to 10\% may have formed in the past few 100\,Myr, with a
clearly increased rate of star formation in the last tens of millions
of years. 

The systematic differences between the stellar evolutionary models
that we have found here
opens up the exciting possibility that we may be able to calibrate
these models observationally in the NSD
if its star formation history can be verified
independently, for example through spectroscopy. Stellar evolution models
are also poorly constrained by observational data at high
metallicities. The Galactic Centre thus offers   an excellent opportunity to
calibrate such models.

The star formation history that we find for the field (i.e.\
non-cluster) population in the Quintuplet field is consistent, within
the uncertainties, with the results of \citet{Nogueras-Lara:2020pp}
and \citet{Nogueras-Lara:2022jz}, but is fully independent from the
latter in important aspects. First, we used a different instrument and
different filters. Second, our dereddening method is similar, but
markedly different in methodological aspects. In particular, we
  exploited the three possible colour combinations from three filters
  instead of a single colour; we did not create an extinction map, but
  determined reddening for each star individually; and we did not apply any
  weighting of the dereddening stars as a function of distance
  \citep[see Methods section in ][]{Nogueras-Lara:2020pp}.  Third, we
fit the cumulative luminosity function, not the binned function, and we
developed new code without recycling   that used in
\citet{Nogueras-Lara:2020pp}. Fourth, we used three different stellar
evolutionary models and a larger number of age bins. We find a
somewhat lower fraction of old stars and a higher fraction of stars
formed in the 1\,Gyr event than \citet{Nogueras-Lara:2020pp}. This
may be due to methodological differences, and also to the fact that the
field studied here is significantly smaller. There are
possbile variations in the population across the NSD, which must be explored by
future work.

A limitation of this work is the relatively bright luminosity cut-off
that we have to use. Therefore our model fitting has to rely on the
locations, shapes, and relative weights of the bright cut-off, the
AGB, the RCB, and the RGBB. Simulations have verified the validity of
this method \citep[see][]{Nogueras-Lara:2020pp,Nogueras-Lara:2022jz}.
Observations at higher angular resolution are needed to push towards
fainter magnitudes, and thus be able to probe the main sequence
turn-off of populations of different ages, which would provide
powerful new constraints on the star formation history.

  Star formation at the Galactic Centre depends upon the
  availability of gas, which is thought to be funnelled towards this
  region by the Galactic bar. Therefore, nuclear stellar discs appear
  to be present in most barred spirals \citep{Gadotti:2020xq}.  Dating
  the formation time of nuclear stellar discs may provide a very good
  proxy to the formation time of Galactic bars
  \citep[see][]{Baba:2019wg}.  Our present analysis suggests an even
  older formation of our Galaxy's nuclear stellar disc than
  \citet{Nogueras-Lara:2020pp}. We speculate that the formation of the
  nuclear stellar disc may coincide with the time when the Milky Way
  underwent its last major merger about 10\,Gyr ago
  \citep{Helmi:2018fz,Kruijssen:2020ev}. \citet{Nogueras-Lara:2020pp}
  pointed out that the 1\,Gyr star formation event may be related to
  the last pericentre passage of the Sgr dwarf galaxy. Related to this
  idea, \citet{Ruiz-Lara:2020ze} suggested that the Sgr dwarf galaxy
  may have induced an episode of enhanced star formation 1\,Gyr ago in
the solar neighbourhood. 

The observation that star formation may have been suppressed between
the original formation of the nuclear stellar disc and the 1\,Gyr
event is puzzling because bars are very efficient at transporting gas
towards galaxy centres. The Milky Way bar is estimated to provide of
the order of 1\,M$_{\odot}$\,yr$^{-1}$ to the Galactic Centre
\citep[e.g.][]{Hatchfield:2021mb}. Bars are generally considered to be
stable long-lived features, and there is reason to believe that the
Milky Way bar is at least 8\,Gyr old \citep[e.g.][]{Bovy:2019az}. An
old formation time of the bar is also evidenced by the age of the
nuclear stellar disc, as discussed above and inferred from work on
Mira variables \citep{Sanders:2022gm}. It therefore appears unlikely
that the Milky Way bar was absent for several
gigayears. 

We can speculate that there was little gas available for star
formation in the inner several
hundred  parsec of the Milky Way during billions of
years. Outflows can remove a considerable fraction of gas from the Galactic Centre
\citep[see discussion in ][]{Henshaw:2022nm}, but either accretion
onto Sgr\,A* or intense star formation are required to drive strong
outflows over long periods of time. Since both require gas inflow as
well, the driving of outflows over billions of years does not appear to be
plausible. Alternatively, gas may have stalled or formed stars at
larger radii (e.g. near the edges of the bar). 

Finally, \citet{Bittner:2020qx} find that nuclear discs in external
galaxies form from the inside out. In this context we note that
in their study of the Sgr\,B1 region, \citet{Nogueras-Lara:2022jz}
report tentative evidence that the outer region of the Milky Way's
nuclear disc contains a significant fraction of stars that formed at
intermediate ages of 2-7\,Gyr. In addition, using proper motion
measurements, \citet{Nogueras-Lara:2023xd} report an inside-out age
gradient for the nuclear stellar disc along the line of sight.  Since
the present work and the previous study by
\citet{Nogueras-Lara:2020pp} are limited to projected distances of
$<50$\,pc from Sgr\,A*, this is a likely explanation for the absence
of intermediate-age stars in these two studies.

Further progress will require us to study the star formation history
in the nuclear disc at larger distances from Sgr\,A* and to use larger
proper motion samples to disentangle the stellar populations. In addition,
deeper and higher angular resolution observations of the stellar
population will be extremely helpful because they could reach the
main sequence turn-off of the oldest populations, and would thus be a highly
reliable age indicator.

\begin{acknowledgements}
  RS, AMA, AG, EGC, MCG, and ATGC  acknowledge financial support from the State Agency for Research  of the Spanish MCIU through the "Center of Excellence Severo Ochoa"
  award for the Instituto de Astrof\'isica de Andaluc\'ia
  (SEV-2017-0709) and financial support from national project
  PGC2018-095049-B-C21 (MCIU/AEI/FEDER, UE). M.H. is supported by the
  Brinson Prize Fellowship. FNL gratefully acknowledges the sponsorship provided by the European Southern Observatory through a research fellowship.
\end{acknowledgements}

\bibliography{/Users/rainer/Documents/BibDesk/BibGC}

\end{document}